# DATA SCIENCE METHODOLOGY FOR CYBERSECURITY PROJECTS

Farhad Foroughi[1] and Peter Luksch[2]

Institute of Computer Science University of Rostock, Rostock, Germany

[1] farhad.foroughi@uni-rostock.de

[2] peter.luksch@uni-rostock.de

## ABSTRACT

*Cybersecurity solutions are traditionally static and signature-based. The traditional solutions along with the use of analytic models, machine learning and big data could be improved by automatically trigger mitigation or provide relevant awareness to control or limit consequences of threats. This kind of intelligent solutions is covered in the context of Data Science for Cybersecurity. Data Science provides a significant role in cybersecurity by utilising the power of data (and big data), high-performance computing and data mining (and machine learning) to protect users against cybercrimes. For this purpose, a successful data science project requires an effective methodology to cover all issues and provide adequate resources. In this paper, we are introducing popular data science methodologies and will compare them in accordance with cybersecurity challenges. A comparison discussion has also delivered to explain methodologies' strengths and weaknesses in case of cybersecurity projects.*

## KEYWORDS

*Cybersecurity, Data Science Methodology, Data-Driven Decision-making, User Data Discovery, KDD Process, CRISP-DM, Foundational Methodology for Data Science, Team Data Science Process*

## 1. INTRODUCTION

Cybersecurity solutions are traditionally static and signature-based which means they depend on pattern identification by detecting a match between a pre-captured attack or a malware with a new threat [1]. Hence, it needs regular update to deploy new signatures in the product database. For this reason, it is not possible to detect or prevent zero-day attacks.

The traditional solutions along with the use of analytic models, machine learning and big data could be improved by automatically trigger mitigation or provide relevant awareness to control or limit consequences of threats. Furthermore, traditional solutions are very binary with limited advantages compared to predictive models that could predict the possibility of attacks or risky actions based on data analysis techniques. In addition, access to a large amount of data makes it possible to solve challenging and complicated security problems. In accordance with big data and data mining, the more data creates more accurate and precise analysis [1].

This kind of intelligent solutions is covered in the context of Data Science for Cybersecurity. A general definition of data science is the information extraction and knowledge discovery from data by using a scientific approach. Data science could build innovative cybersecurity solutions by utilising new technological advantages of storage, computing and behavioural analytics [2]. In general, cluster storages which are deployed by distributed systems make it easier to collect and store huge amount of data (big data), and cloud computing also make it possible to utilise complex and sophisticated machine learning techniques to create predictive and analytic models to identify and detect and respond threats. Behavioural analytics also could transform traditional signature-based detection techniques to the new behaviour-based predictive solutions.

According to data analysis potentials in cybersecurity, National Institute of Standards and Technology has developed a framework consists asset risk identification (and threat consequences), information protection, intruders detection, responding to intruders and business recovery [3].

As a result, Data Science provides significant role in cybersecurity by utilising the power of data (and big data), high-performance computing and data mining (and machine learning) to protect users against cybercrimes. For this purpose, a successful data science project requires an effective methodology to cover all issues and provide adequate resources.

In this paper, we are introducing popular data science methodologies and will compare them in accordance with cybersecurity challenges. Section 2 describes general Data Science overview along with its relation to the cybersecurity concept. Section 3 provides information about popular data science methodologies in details. Four different methodologies have been explained in this section. A comparison discussion is delivered in section 4 to explain methodologies' strengths and weaknesses against each other along with a summary table. In the end, we recommend a methodology that might cover all requirements to provide the most possible efficient data science cybersecurity project.

## 2. DATA SCIENCE

Data science could enhance and improve decision-making process by providing data-driven predictions. This requires principles, processes and techniques to understand a problem through an automated evaluation and data analysis. A successful data science has to employ data mining algorithms to solve business problems from a data perspective [4].

Data science is a set of basic concept which leads to the principled extraction of information and knowledge from data. It is very similar to the data mining tries to extract this information via technologies and applied and utilised for relationship management and behaviour analysis in order to recognize patterns, values and user interests [4]. CFJ Wu identified differences between traditional pure statistics and modern data science practices in 1997. He described these significant factors as Data Collection, Data Modelling and Analysis, and Problem Solving and Decision Support [5]. Data science is a recursive process which requires iterative performing.

Fayyad says Data Mining is a component of knowledge discovery in database process [6]. The knowledge discovery in database (KDD) was composed in 1989 to refer to the wide practice of obtaining and acquiring knowledge in data to stress the high-level application of certain data mining techniques. According to Fayyad et al. definition, KDD is the process of utilising data mining techniques to draw out knowledge based on the specification of measures and threshold, making use of a database together with any necessary pre-processing, sampling and transforming the database [7]. Therefore, a knowledge discovery process requires at least, Selection, Pre-processing, Transformation, Data mining, Interpretation and Evaluation steps. Knowledge discovery in database in cybersecurity domain interpreted into two major concepts. These concepts are User Data Discovery (UDD) with is a user profiling process and Data-Driven Decision-making which is a decision-making process based on data analysis [8].

## 2.1. Data-Driven Decision-making

Data-driven decision-making (DDD) is the term for the process and technique of taking decisions based on data analysis and information evaluation rather than strictly on intuition [4].

DDD is not a binary practice to provide all or nothing. It could be employed in cybersecurity domain with different levels of engagement. Provost et al. demonstrate two types of decisions: 1) decisions which are based on data discovery 2) decisions which are based on frequent decision-making processes particularly at considerable dimensions or massive scale. This kind of decision-making processes might gain from an even minor increase in reliability and precision based on information evaluation and data analysis [4].

Figure 1 describes data science and data-driven decision-making relation. Data Science overlaps data-driven decision-making because cybersecurity decisions and choices could increasingly be created instantly and automatically by computer systems [4].

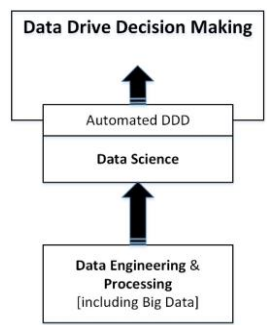

Figure (1): Data-Drive Decision-making through Data Science

Data processing and data engineering are essential to support data science tasks and very beneficial for data-driven decision-making, effective transaction processing and online pattern recognition. Big Data is simply a term for datasets which are very large for conventional data processing and require new methods and technologies. Therefore, big data technologies are in fact utilised for applying data processing to support data mining strategies and data-driven decision-making tasks [4]. Modern efficient cybersecurity solutions are depend on big data because more data creates more accurate and precise analysis [1].

Data analytic thinking is a crucial element of data science. Underlying the comprehensive collection of methods and strategies for mining information is a significantly smaller set of basic concepts comprising data science. Understanding the essential concepts and having a data analytic thinking structure and framework could help cybersecurity researchers to boost data-driven decision-making process.

## 2.2. User Data Discovery

User Data Discovery (UDD) is the process of producing profile of users from historical information and past details. This particular information might be personal data, academic documents, geographical details or other private activities.

The primary function of user profiling process is capturing user's information about interest domain. These information may be used to understand more about individual's knowledge and skills and to improve user satisfaction or help to make a proper decision. Typically, it evolves data mining techniques and machine learning strategies. UDD process is a type of knowledge discovery in database or the new version, knowledge data discovery model and requires similar steps to be established [8].

User profiling is usually either knowledge-based or behaviour based. The knowledge-based strategy uses statistical models to categorise a user in the closest model based on dynamic attributes. Typically, Questionnaires and interviews could be utilised to acquire this particular user knowledge [9].

Behaviour-based strategy employs the user's behaviours and actions as a model to observe beneficial patterns by applying machine learning techniques. These behaviours could be extracted through monitoring and logging tasks [9].

Recognizing user behaviour in real time is an important element of providing appropriate information and help to take suitable action or decision in cybersecurity projects. Typically it is a human task that experts would provide in the information security domain, but it is possible to employ user modelling to make this process automatic by using an application or intelligent agent [10].

UDD could obtain appropriate, adequate and accurate information about a user's interests and characteristics and demonstrate them with minimal user intervention [11] to offer appropriate awareness with relevant mitigation recommendation based on the security situation. An intelligent cybersecurity solution should take into account the various attributes and features of a user and a security situation to create a customized solution based on the notion and concept of user profile [12].

## 3. DATA SCIENCE METHODOLOGY

Several theoretical and empirical researchers have considered basic concepts and principles of knowledge extraction from data. These basic methods and fundamental principles are concluded from numerous data analytic studies [4].

Extracting beneficial knowledge from data should be dealt with systematic processes and procedures through well-defined steps.

Data science needs attentive consideration and result evaluation in the context it is used because the extracted knowledge is significant to assist the decision-making process in a particular application [4].

"Breaking the business problem up into components corresponding to estimating probabilities and computing or estimating values, along with a structure for recombining the components, is broadly useful." [4]

The correlation finding is one of the data science concepts which should be considered in relation to the cybersecurity. It typically provides details on data items that supply information about other data items, particularly, recognized quantities which reduce the uncertainty of unknown quantities [4].

Entities which are identical with regard to known features or attributes, oftentimes are identical with regard to unknown features or attributes. Computing similarity (pattern recognition) is among the primary resources of data science [4]. It is also significant to pay quite close attention to the existence of confounding elements, most likely unseen ones.

A methodology is a general approach that guides the techniques and activities within a specific domain. The methodology does not rely on certain technologies or tools. Instead, a methodology delivers a framework to acquire results by using a wide range of methods, processes and heuristics [13].

Predictive model creation, pattern recognition and underlying discovery problems through data analysis are usually a standard practice. Data science provides plenty of evolving data analysis technologies to constructing these models. Emerging analytics methods and action automation provide strong machine learning models to solve sophisticated analytic problems such as DDDs. To create an appropriate data analytic model it is required to use a data science methodology that could provide and supply a guiding strategy regardless of technology, data volumes or approaches.

There are several methodologies available for data mining and data science problems such as CRISP-DM and SAS SEMMA and KDD process but Gregory Piatetsky confirms CRISP-DM remains the top methodology for data mining projects with 42% in 2014. The KDD Process has used by 7.5% [14].

Rollins demonstrates a foundational methodology which is similar to CRISP-DM, SEMMA and KDD Process but also emphasizes a number of new methods in data science including big data usage, the incorporation of text analytics into predictive modelling and process automation [13]. Microsoft also introduces Team Data Science Process (TDSP) which recommends a lifecycle for data science projects [15].

Before applying any of these methodologies to cybersecurity projects, it might be helpful to review and compare their essential features. For this reason, this paper provides a comparison between KDD Process, CRISP-DM, TDSP and the foundational methodology for data science (FMDS). FMDS and CRISP-DM have been chosen, because they are considered to be the most popular but SAS SEMMA is not in this review because there is a big decline in applying it (from 13% in 2007 to 8.5% in 2014) [14]. The KDD Process has also included because it provides initial and basic requirements of knowledge discovery. TDSP has been chosen because it is customized for machine learning or artificial intelligence projects which are considerably linked to cybersecurity applications.

### 3.1. KDD Process

The KDD process offered by Fayyad et al. in 1996 [7]. It is the method of using data mining techniques to extract knowledge based on particular measures and thresholds in a database by employing any necessary pre-processing, sampling or data transformation actions [7]. Furthermore, the application domain perception is needed to be considered during the KDD process development, improvement and enhancement. Figure 2 illustrates the KDD process.

The KDD process has 5 steps as the following [7]:

1. Selection: It means generating or producing a target data set or concentrating on a subset of variables or data samples in a database.

2. Pre-processing: This phase tries to obtain consistent data by cleaning or pre-processing selected data.

3. Transformation: Reducing feature dimensionality by using data transformation methods is performing in this phase.

4. Data Mining: Trying to recognize patterns of attention or behaviours by using data mining techniques in a specific form should perform in this step. (Typically, prediction)

5. Interpretation/Evaluation: Mined pattern should be assessed and interpreted in the final phase.

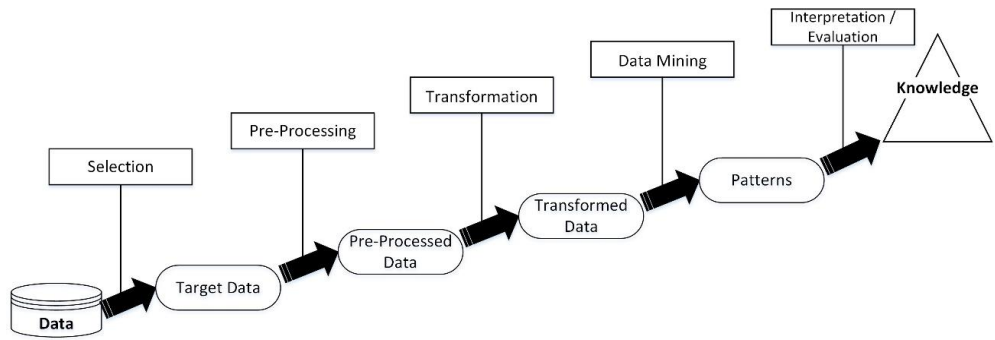

Figure (2): The KDD Process

## 3.2. CRISP-DM

In 1996, SPSS and Teradata developed Cross Industry Standard Process for Data Mining (CRISP-DM) in an effort initially composed with NCR, Daimler and OHRA. It is a cycle of six high-level phases which describe the analytic process [16, 17].

CRISP-DM is still a beneficial tool but details and specifics needed to be updated for cybersecurity projects such as those including Big Data. The original site is not active anymore but IBM SPSS modeller still supports it [14].

Figure 3 demonstrates the CRISP-DM six stages, but their sequence is not strict. CRISP-DM is very well documented and there are many case studies which have used it [16].

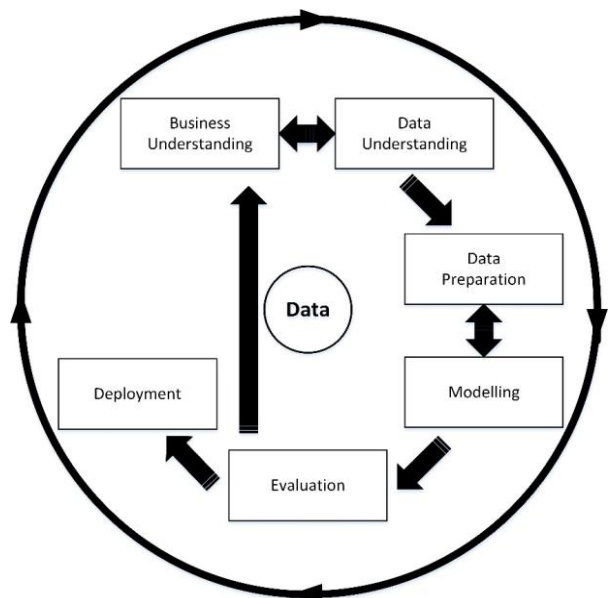

Figure (3): The CRISP-DM life cycle

CRISP-DM's structured, well defined and extremely documented process is independent of data mining tools and this important factor is crucial to make the project successful [16].

The CRISP-DM is so logical and seems like common sense. There are many methodologies and advanced analytic platforms which are actually based on CRISP-DM steps because the use of a commonly practised methodology gains quality and efficiency. Vorhies says CRISP-DM provides strong guidance for even the most advanced of today's data science projects [17].

The six phases are the following [16]:

1. Business understanding: It is designed to focus on understanding the problem or project goals and requirements from a business perspective (here is a cybersecurity application) and then transforming this perception into data mining problem description and preliminary approach.

2. Data Understanding: This phase begins with an initial data collection and then with tasks in order to acquaint with data, to recognise data quality problems, to find out primarily insights into the details or even to identify interesting subsets to develop hypotheses for hidden information. It typically creates a set of raw data.

3. Data Preparation: This phase covers all tasks and activities to build the final required dataset from the first raw data.

4. Modelling: Through this stage, modelling techniques and strategies are selected and applied, and their specific parameters and prerequisites should be identified and calibrated regarding the type of data to optimal values.

5. Evaluation: At this point, the obtained model or models which seem to provide high quality based on loss function completely evaluated and the actions executed to ensure they generalise against hidden data and to be certain it correctly archives the key business goals. The final result is the selection of sufficient model or models.

6. Deployment: This stage means deploying a code representation of the final model or models in order to evaluate or even categorise new data as it arises to generate a mechanism for the use of new data in the formula of the first problem. Even if the goal of the model is to provide knowledge or to understand the data, the knowledge acquired have to be organised, structured and presented in a means which could be used. This includes all the data preparation and required steps which are needed to treat raw data to achieve the final result in the same way as developed during model construction.

### 3.3. Foundational Methodology for Data Science

This methodology has some similarities and consists many features of KDD Process and CRISP-DM, but in addition, it provides a number of new practices such as use of extremely large data volumes, text and image analytics, deep learning, artificial intelligence and language processing [13]. The FMDS's ten steps illustrate an iterative nature of the problem-solving process for utilising data to discover security insights. Figure 4 demonstrates FMDS process.

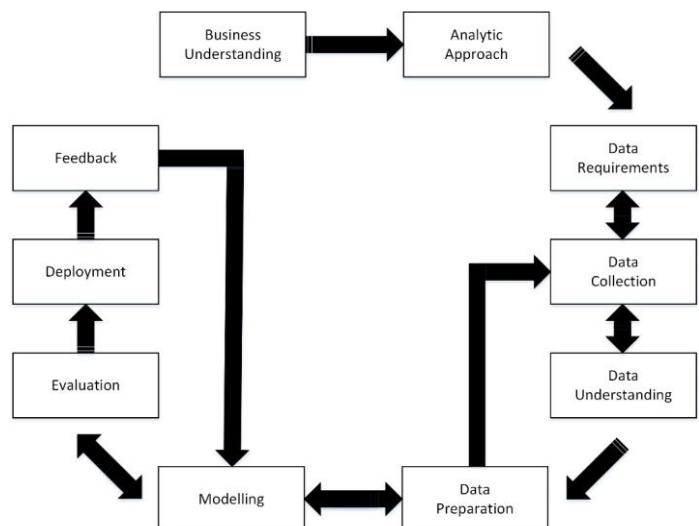

Figure (4): Foundational Methodology for Data Science

Foundational Methodology for Data Science consists the following steps [18]:

1. Business understanding: This very first phase provides a basic foundation for a profitable and effective resolution of a business problem (a cybersecurity challenge in here) regardless of its size and complexity.

2. Analytic approach: As soon as the problem clearly stated, it is required to determine an analytic approach by identifying suitable machine learning technique to solve it and obtain the desired result.

3. Data requirements: The analytic approach that has been chosen in the second phase, defines needed data requirements including specific data content, formats and representations which are instructed by cybersecurity knowledge.

4. Data collection: In this primary data gathering phase, it is required to identify and collect available data resources (structured, unstructured and semi-structured) which are related and applicable to the problem domain. It is important to cover any data collection gap by revising data requirements and gathering brand new and/or additional data.

It is also significant to use high-performance platforms and in-database analytic functionality to work with huge datasets for sampling and sub-setting to obtain all available data.

5. Data understanding: Descriptive statistics and visualisation methods are useful in this phase to understand data content, evaluate data quality and explore data insights. This could be required to revise the earlier phase to close data collection gaps.

6. Data preparation: This phase comprises all tasks to construct dataset which will be utilised in the modelling phase. These tasks include data cleaning, data merging from several sources, dealing with missing data, data transformation into more useful variables, duplication elimination, and finding invalid values. In addition, feature engineering and text analytics are possible to be utilised to provide new structured variables, defining and enriching the predictors and boosting or improving the model's reliability, accuracy and precision. A collaboration of cybersecurity knowledge and existing structured variables could be very useful for feature engineering. This phase is probably the most time-consuming stage, but high-performance and parallel computing systems could reduce the time required and prepare data quickly from huge datasets.

7. Modelling: This phase concentrate on predictive or descriptive model development based on earlier described analytic approach and by using the first version of the prepared dataset as a training set (historical data). The modelling process is extremely iterative as it provides intermediate insights and reputable refinement of data preparation and model specification. It is significantly helpful to try several algorithms with specific parameters to find the ideal model.

8. Evaluation: Before deployment, it is crucial to evaluate the quality and efficacy of the developed model to realise whether it completely and appropriately addresses the cybersecurity problem. This evaluation involves computing of several different diagnostic measures and other outputs including tables and graphs by using a testing set. This testing set is actually independent of the training set but follows the identical probability distribution and has known results.

9. Deployment: Once the developed model approved and accredited in the evaluation phase that covers the cybersecurity challenge appropriately, it should be deployed into the production environment or perhaps in a comparable test environment. Typically, it will be used in a limited and specific way until all performance variables entirely evaluated. Deployment could be as basic as producing a simple report with proper suggestions or perhaps, embedding the model in an elaborated or sophisticated workflow and scoring process handled by a customised application.

10. Feedback: This final phase, collects outcomes from the implemented edition of the analytic model to analyse and feedback its functionality, performance and efficiency in accordance with the deployment environment.

### 3.4. Team Data Science Process

The Team Data Science Process (TDSP) is a data science methodology to provide efficient predictive analytics. TDSP These solutions are including artificial intelligence and machine learning. It boots data science project agility, team working and learning by employing best practices and successful structures from Microsoft [15]. TDSP supports both exploratory and ad-hoc projects. Figure 5 illustrates TDSP five stages.

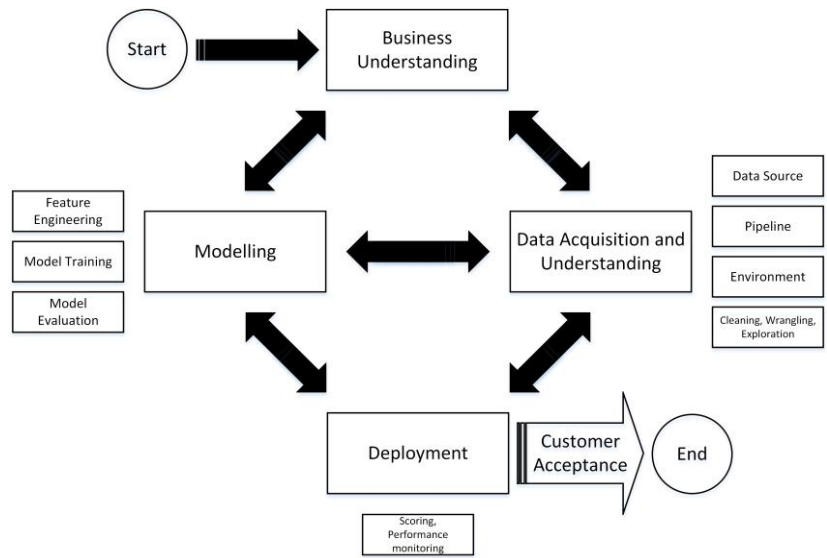

Figure (5): Team Data Science Process Lifecycle

TDSP supports development of projects which have already employed CRISP-DM and KDD process. It is very customizable based on project's size and dimensions [19].

The TDSP lifecycle consists integrative phases as the following [19]:

1. Business Understanding: Initially, a question which describes the problem objectives should be defined clearly and explicitly. Relevant predictive model and required data source/s also have to be identified in this step.

2. Data Acquisition and Understanding: Data collection starts in this phase by transferring data into the target location to be utilised by analytic operations. The raw data is needed to be cleaned and incomplete to incorrect values should be identified. Data summarization and visualization might help to find required cleaning procedures. Data visualization also could help to measure if data features and collected amount of data are adequate over the time period. At the end of this stage, it might be necessary to go back to the first step for more data collection.

3. Modelling: Feature engineering and model training are two elements of this phase. Feature engineering provides attributes and data features which are required for machine learning algorithm. Algorithm selection, model creation and predictive model evaluation are also sub-components of this step. Collected data should be divided into training and testing datasets to train and evaluate machine learning model. It is important to employ different algorithms and parameters to find the best suitable solution to support the problem.

4. Deployment: Predictive model and data pipeline are needed to be produced in this step. It could be a real-time or a batch analysis model depends on the required application. The final data product should be accredited by the customer.

5. Customer Acceptance: The final phase is customer acceptance which should be performed by confirming data pipeline, predictive model and product deployment.

## 4. DISCUSSION

All data science methodologies where discussed consist four common iterative stages including problem definition/ formulation, data gathering, data modelling and data product development except the KDD process.

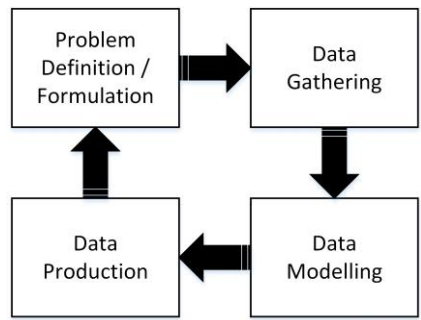

Figure (6): General Data Science Methodology

According to the above explanations, table 1 demonstrates a summary of the presented correspondences.

Table (1): Summary of Data Science methodologies and their phases

| KDD Process | CRISP-DM | FMDS | TDSP |
|---|---|---|---|
| - | Business Understanding | Business Understanding | Business Understanding |
| - | - | Analytic Approach | |
| - | - | Data Requirements | |
| Selection | Data Understanding | Data Collection | Data Acquisition and Understanding |
| Pre Processing | | Data Understanding | |
| Transformation | Data Preparation | Data Preparation | |
| Data Mining | Modelling | Modelling | Modelling |
| Interpretation/ Evaluation | Evaluation | Evaluation | |
| - | Deployment | Deployment | Deployment |
| - | - | Feedback | Customer Acceptance |

Comparing the KDD process with the CRISP-DM presents that KDD process does not cover the business understanding and also deployment phases in CRISP-DM methodology.

As it is mentioned above, the business understanding phase provides a comprehension perception of the application domain and pertinent prior knowledge and also objectives of required solution. Furthermore, the deployment phase incorporates knowledge or modelling code into a system or application to build a data product. These two significant phases have been missed in KDD process.

Comparing the CRISP-DM with the FMDS illustrates that CRISP-DM has not the analytic approach and the feedback phases. The analytic approach is required to recognize appropriate statistical or machine learning techniques before entering to the data gathering steps. This phase could be very useful to identify the suitable data collection strategy and data resources but it is missed in the CRISP-DM methodology. The feedback phase also has been missed that is very beneficial to optimize the system to achieve high-performance functionality and efficient result.

A comparison between FMDS and TDSP presents that they are very similar but FMDS has more details steps in general. Feedback also is a part of FMDS cycle which could create new requirements to improve the data product in an iterative process but customer acceptance is a stage out of the circle in the TDSP. FMDS detailed stages could be more useful for a wide range of projects but TDSP uses a specific set of Microsoft tools and infrastructure to deliver intelligent applications by deploying machine learning or AI models.

Concerning the remaining phases present the following:

- The Data understanding phase is in both CRSIP-DM and FMDS can be recognized as the collaboration of Selection (Collection) and Pre-Processing phases in the KDD process but Data Acquisition and Understanding in TDSP also covers Transformation stage in the KDD process. However, data requirements which are related to the analytic approach phase and provides required data content are missed in both KDD process and CRISP-DM. Selecting an analytic approach is integrated into the business understanding phase of TDSP but cybersecurity projects might gain more benefits in details from this task when it is an independent step in the FMDS particularly when data resources are separated and requires a different method or level of access.

- Business understanding and problem formulation is an initial phase that makes the data understanding phase more efficient by recommending data formats and representations but data requirement analysis is missed in the CRISP-DM. It might be very crucial in cybersecurity projects particularly when data resources are unstructured or semi-structured.

- The data preparation phase has the similar function as the transformation phase in KDD process and it is included in the Data Acquisition and Understanding phase of TDSP by using some tools.

- The modelling phase might be recognised with data mining phase which is very limited in the KDD process. Modelling phase in TDSP also has the evaluation task included by using some tools such as scoring and performance monitoring, but it is an independent phase in other methodologies.

- The evaluation phase is also included in all three methodologies.

In spite of CRISP-DM strengths, there is more emphasis that should be considered for modern cybersecurity projects.

- It is required that methodology could also handle data blending from several sources. This should be deployed through a completely repeatable process. FMDS and TDSP provide this feature in the Data Preparation and Data Acquisition and Understanding phases. TDSP provides Microsoft tools to support On-Premises, Cloud, Databases and Files but FDMS is independent of any platform or a specific tool. This might makes cybersecurity projects more reliable and efficient.

- Choosing the most appropriate degree of reliability and accuracy for the problem is very crucial to make sure there is no need to spend excessive time on data preparation or modelling to enhance accuracy when it could not be utilised. This feature also included in the FMDS methodology.

- The entire analytic algorithm should be tested and evaluated to make sure there are working in all situations and not just for sample modeller. The evaluation, deployment and feedback cycle in the FMDS could provide this requirement as well as Model training task of Modelling phase in the TDSP. Feedback phase in FMDS might create new data science questions to optimize the cybersecurity product or make new functionalities for it.

- It is significant to consider quality during model simplification by ensuring that decision elements such as missing data reduction, synthetic features generation and unseen data holdout are properly managed. The evaluation, deployment and feedback cycle in the FMDS could bring this need better than simple quality insurance in the evaluation phase of CRISP-DM.

- Data science lifecycle is very well defined in the FMDS and connections are clearly determined between every stage but TDSP's stages are all linked together (except customer acceptance) and it is possible to move into any stages from anyone else.

- TDSP lifecycle is designed for predictive analytic projects by using machine learning or artificial intelligence models. FMDS is more general and independent of any platform, tool, model or algorithm. Both are functional for exploratory or ad-hoc cybersecurity analytics projects by some customization.

## 5. CONCLUSION

In conclusion, a cybersecurity data science project requires four general steps. The first step has to be a problem definition by formulating a security challenge. In accordance with problem definition and appropriate formula, it is necessary to gather required information in the second step. The collected information should be employed in the third step and in an analysis process to provide adequate data which is expected to predict or provide a resolution for the defined problem. The final step is a production step which deploys relevant modules and a system to run the whole process automatically and regularly when it is needed.

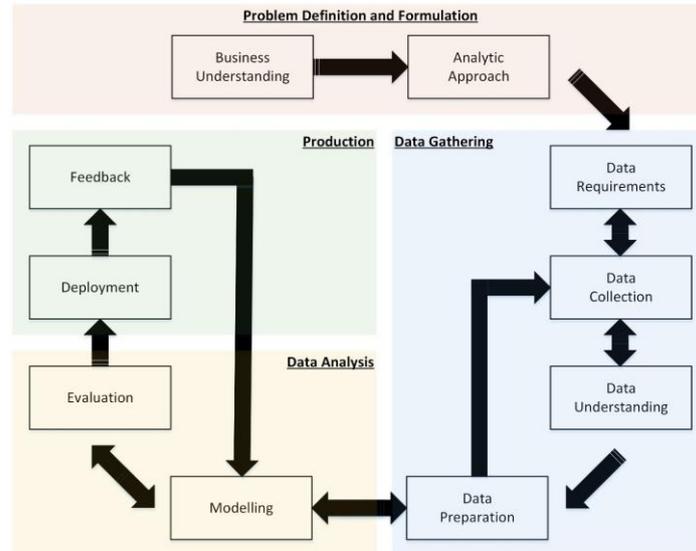

Figure (7): FMDS for Cybersecurity projects

Regarding the general process and in accordance to Table 1, the FMDS covers all beneficial attributes of CRISP-DM methodology but fills data gathering gaps and also provides extra steps to optimize and enhance the model and results by mathematical prescriptive analytics and using high-performance computing. It is also platform and tool independent but TDSP is not. Because it is designed in details with clearly divided steps, it could be fully customized to fit in any cybersecurity project.